\documentclass[12pt]{article}
\pdfoutput=1
\usepackage{hyperref}
\usepackage{cite}
\usepackage{xcolor}
\usepackage{graphicx}
\usepackage{caption,subcaption}
\usepackage{amsmath}
\usepackage{amssymb}
\usepackage{xspace}
\usepackage{verbatim}
\usepackage{mathtools}
\usepackage{tikz-cd}
\usepackage[normalem]{ulem}

\makeatletter
\@addtoreset{equation}{section}

\makeatletter
\renewcommand\section{\@startsection {section}{1}{\z@}%
                                   {-3.5ex \@plus -1ex \@minus -.2ex}
                                   {2.3ex \@plus.2ex}%
                                   {\normalfont\large\bfseries}}
\renewcommand\subsection{\@startsection{subsection}{2}{\z@}%
                                     {-3.25ex\@plus -1ex \@minus -.2ex}%
                                     {1.5ex \@plus .2ex}%
                                     {\normalfont\bfseries}}

\def\baselinestretch{1.2}
\newcommand{\be}{\begin{equation}}
\newcommand{\ee}{\end{equation}}
\newcommand{\beq}{\begin{eqnarray}}
\newcommand{\eeq}{\end{eqnarray}}

\newcommand{\gone}[1]{{}}

\begin{document}
\begin{titlepage}
\begin{flushright}
MAD-TH-17-06
\end{flushright}

\vfil

\begin{center}

{\bf \Large
Gravitational Decoupling and Picard-Lefschetz
}

\vfil

Jon Brown, Alex Cole, William Cottrell, and Gary Shiu

\vfil

Department of Physics, University of Wisconsin, Madison, WI 53706, USA

\vfil
\end{center}

\begin{abstract}

\noindent

In this work, we consider tunneling between non-metastable states in gravitational theories.  Such processes arise in various contexts, e.g., in inflationary scenarios where the inflaton potential involves multiple fields or multiple branches.  They are also relevant for bubble wall nucleation in some cosmological settings.  However, we show that the transition amplitudes computed using the Euclidean method generally do not approach the corresponding field theory limit as $M_{p}\rightarrow \infty$.  This implies that in the Euclidean framework, there is no systematic expansion in powers of $G_{N}$ for such processes.  Such considerations also carry over directly to no-boundary scenarios involving Hawking-Turok instantons.  In this note, we illustrate this failure of decoupling in the Euclidean approach with a simple model of axion monodromy and then argue that the situation can be remedied with a Lorentzian prescription such as the Picard-Lefschetz theory. As a proof of concept, we illustrate with a simple model how tunneling transition amplitudes can be calculated using the Picard-Lefschetz approach.

\end{abstract}
\vspace{0.5in}

\end{titlepage}
\renewcommand{\baselinestretch}{1.05}  

\section{Introduction}

Euclideanization is a technique that is ubiquitous throughout physics.  It is used to understand problems including the origin of the universe, numerical computations in field theory, non-perturbative transitions and instantons, perturbative evaluation of the path integral, etc.  Its power as a computational tool stems from its effect on the path integral, transforming an oscillatory integral in Lorentzian signature into an exponentially suppressed integral in Euclidean signature
\begin{equation}
	\int\mathcal{D}\phi \; e^{iS\left[\phi\right]} \to \int\mathcal{D}\phi \; e^{-S_E\left[\phi\right]}
\end{equation}

Despite these successes, Euclidean techniques are not without limitations.  For example, treating a system of fermions at nonzero temperature and chemical potential can give the action a nonzero imaginary part making the path integral oscillatory in either signature, a problem commonly encountered when putting QCD on a lattice~\cite{deForcrand:2010ys}.  In gravitational theories the Euclidean action is not semipositive definite, allowing exponential enhancement rather than suppression in the Euclidean path integral for certain deformations~\cite{Gibbons:1978ac}.

An additional limitation of Euclidean techniques is encountered in dynamical situations, in particular, quantum tunneling in dynamical backgrounds.  Consider, for instance, a pair of scalar fields in the potential described by
\begin{equation}
	\label{NewEffPot}
	V\left(\phi,q\right) = \frac{1}{2}\mu^2\left(\phi + \frac{q}{\mu}\right)^2 + V_\text{UV}\left(q\right)
\end{equation}
Here, $\phi$ is an axion field and $q$ is a second field that is tightly constrained around the minima of $V_{\text{UV}}(q)$, which are assumed to occur at evenly spaced intervals, $q = n e$.  This is the kind of potential relevant for tunneling in axion monodromy models of inflation \cite{Silverstein:2008sg,McAllister:2008hb,Kaloper:2008fb,Kaloper:2011jz,Marchesano:2014mla,Blumenhagen:2014gta,Hebecker:2014eua} and is also of interest for recent discussion of bubble wall nucleation in the early universe~\cite{Deng:2016vzb,Garriga:2015fdk}.  In~\cite{Brown:2016nqt}, three of the current authors studied field-theoretic tunneling in this potential using the Euclidean prescription.  The motivation of this previous work was to investigate tunneling through the barrier of the $V_{\text{UV}}$ potential {\it{while}} the axion field $\phi$ is rolling down its quadratic potential.   This is a natural extension of the kind of vacuum transition studied by Coleman~\cite{Coleman:1977py}.

However, various issues arise when applying Euclidean methods to generic multi-field tunneling.  For instance, the primary justification for Euclideanization in this context is based on the WKB approximation, which, at fixed energy, approximates the wavefunction with a local plane wave, $\psi\left(x\right) \sim e^{i \int p\left(x\right) dx}$, where the local momentum is $p\left(x\right) \sim \sqrt{2m\left(E - V\left(x\right)\right)}$.  One can see that the momentum of a particle becomes imaginary when passing underneath a potential barrier and this can be equivalently described by instead Euclideanizing the time coordinate~\cite{Goncharov:1986ua}.  However, if we have multiple spatial directions in quantum mechanics, multiple fields in QFT, or are not starting in an energy eigenstate, WKB and Euclideanization do not automatically agree, thus calling the latter procedure into question.  Nevertheless, many attempts have been made~\cite{Banks:1973ps, Banks:1974ij,Bowcock:1991dr, Widrow:1991xu} to make sense of these kinds of tunneling processes and it is generally believed that the Euclidean prescription provides a good leading approximation.

Let us momentarily adopt the attitude that the Euclidean prescription gives the leading contribution to a full result, despite the lack of  understanding for how to systematically compute subleading corrections to the WKB approximation in a dynamical situation.  In this case, it is still natural to ask about the size of leading corrections coming from a coupling to gravity.  In the context of thin-wall tunneling of a single field, this was addressed by Coleman and De Luccia (CdL)~\cite{Coleman:1980aw}, where it was shown that the gravitational corrections are suppressed by a factor of $\sigma^{2}/\left(|\Delta V| M_{p}^{2}\right)$, where $\sigma$ is the domain wall tension and $\Delta V$ is the jump in the potential during tunneling.  As one might expect, this vanishes in the `decoupling limit', $M_{p}\rightarrow \infty$, where we should naively reproduce the field theory result.

In contrast, we will show in this paper that for the two field potential~\eqref{NewEffPot} and in general other tunneling while rolling situations, one does not approach the field theoretic result as the Planck mass is taken to infinity.  In fact, once the coupling to gravity is considered, one needs UV input from quantum gravity to even define the Euclidean tunneling amplitude associated with~\eqref{NewEffPot}.  This is a highly undesirable situation, as the real world contains gravity, and we would like to consider decoupled field-theoretic systems at low energy without resorting to a full-fledged theory of quantum gravity.  This suggests that the Euclidean result is not even a first approximation in some more systematic approach --  it is instead just the wrong starting point for this kind of problem.

In fact, we will argue that this problem of non-decoupling is fairly generic for Euclidean tunneling computations, especially during inflation.  This is simply because Euclidean solutions with an inflaton field have geometries with singular asymptotics.  In fact, the singularity, first described by Hawking and Turok (HT)~\cite{Hawking:1998bn} is universal in the sense that it emerges for any unbounded inflaton potential\footnote{One must only require that the potential is bounded by a certain exponentially growing function.} and is guaranteed given generic initial conditions.  Thus, to achieve a reasonable solution, one must either introduce a new high-scale metastable vacuum by hand or regularize the singularity somehow.  Either route leads to a strong dependence on UV physics, regardless of the energy scales involved in the original tunneling problem.  These considerations naturally carry over to the no-boundary proposal, where our work suggests that the wavefunction is UV sensitive and not determined by the low energy effective action.

A very similar discussion has unfolded in the recent literature where the authors~\cite{Feldbrugge:2017kzv,Feldbrugge:2017fcc,Feldbrugge:2017mbc} have claimed that the Hartle-Hawking wavefunction~\cite{Hartle:1983ai} is not smooth due to unsuppressed fluctuations in the Euclidean action, while~\cite{DiazDorronsoro:2017hti} has countered that a proper choice of path-integration contour eliminates these issues.  Our point is slightly orthogonal.  While these authors are studying classically smooth geometries and their focus is on the suppression of angular fluctuations, we are discussing singular (yet generic) geometries where the problem is simply a lack of a decoupling limit, even for spherically symmetric perturbations.

In these examples we are forced to consider alternative methods for computing the tunneling amplitude.  We will argue that the Lorentzian path integral does not suffer from a failure to decouple, which is very much in the spirit of~\cite{Feldbrugge:2017kzv} where Lorentzian methods were advocated for slightly different reasons.  Moreover, this approach can potentially be made practical with the help of Picard-Lefschetz (PL) theory~\cite{Witten:2010cx}, which instructs one to complexify fields rather than time.  The net result is again a  manifestly exponentially suppressed path integral, but now one that is directly related to the original Lorentzian path integral via analytic continuation.  In contrast, there is no obvious way to connect the Euclidean path integral to the Lorentzian one via analytic continuation for general boundary conditions.  In this paper, we will provide some examples of how decoupling is successfully achieved within the Picard-Lefschetz framework and also discuss the relationship with the Euclidean theory.

The paper is organized as follows.  In Section~\ref{gd} we first review how gravitational decoupling works in general, in particular reviewing the role of boundary terms.  Next, in Section~\ref{ep} we discuss Euclidean tunneling processes with non-standard initial conditions and show that decoupling fails.  Section~\ref{pl} describes how to circumvent this issue with Picard-Lefschetz.  Finally, we offer our conclusions.

\section{Gravitational decoupling}
\label{gd}

Standard intuition tells us that as the Planck scale is taken to infinity one should approach a field theory limit where gravity can be ignored.  We would like to review how this works in detail.  Schematically, take a gravitational theory coupled to matter

\begin{equation}
	\label{gAction}
	S_g = \int d^4x\sqrt{-g}\left(\frac{M_p^2}{2}\mathcal{R} + \mathcal{L}_\text{matter}(\phi)\right) + S_\text{boundary}
\end{equation}
where the subscript on $S_{g}$ is for `gravitational'.  A successful decoupling of gravity and matter as $M_{p}\rightarrow \infty$\footnote{In taking this limit, we are implicitly keeping energy scales associated with the matter sector fixed.} would mean that the action reduces to that of the decoupled matter, up to a possible constant term
\begin{equation}
	\label{dAction}
	S_g \rightarrow \int d^4x \sqrt{-g_{0}}\, \mathcal{L}_\text{matter}(\phi) +\text{const}
\end{equation}
where $g_{0}$ is some fixed background metric, and we are allowing for off-shell configurations of the matter field $\phi$.  We will be particularly interested in situations where the spacetime is locally flat in some region where the matter fields reside, but may nevertheless have non-trivial asymptotics affecting the constant piece.

Naively, it is difficult to see how~\eqref{dAction} could be achieved since Einstein's equation, $M_p^2G_{\mu\nu} = T_{\mu\nu}$, tells us that the gravitational and matter parts of the bulk action~\eqref{gAction} are always of the same order.  For example, a single canonically normalized scalar with action $\mathcal{L}_\text{matter} = -\frac{1}{2}\left(\partial\phi\right)^2 - V\left(\phi\right)$ leads to the equation
\begin{equation}
	\label{gravityEOM}
	\frac{M_p^2}{2}\mathcal{R} = \frac{1}{2}\left(\partial\phi\right)^2 + 2V\left(\phi\right)
\end{equation}
Thus the Einstein-Hilbert term always contributes at the same order as the matter Lagrangian, regardless of the value of $M_{p}$.  Going one step further, one can plug the equation of motion~\eqref{gravityEOM} back into the action $S_g$ to get
\begin{equation}
	S_g = \int d^4x\sqrt{-g}\left(+V\left(\phi\right)\right) + S_\text{boundary}
\end{equation}
We thus see that the sign of the potential apparently flips.

It is clear from this discussion that if decoupling is to work,  then the boundary contribution $S_\text{boundary}$, as well as the boundary conditions must play a role.  Here, for reasons that will become clear later, we will consider the standard Gibbons-Hawking-York boundary term multiplied by an arbitrary constant 
\begin{equation}
	S_\text{boundary} \to S_{GHY} = C_{GHY} M_p^2\int_{\partial\mathcal{M}} d^3x \sqrt{-h} \left(\mathcal{K} - \mathcal{K}_0\right)
\end{equation}
where $\mathcal{K}$ is the trace of the extrinsic curvature on the boundary manifold $\partial\mathcal{M}$ and $h$ is the induced metric on the boundary.  

Consider now a configuration $(\phi, g)$ that, at least in some large region of interest, is  `close'\footnote{By `close', we mean here that the $|\phi-\phi_{0}|$ as well as the variables appearing in the potential are held fixed as $M_{p}\rightarrow \infty$. Furthermore, we should require that the potential difference $V(\phi)-V(\phi_{0})$ is also bounded uniformly, such that $\Delta V/M_{p}^{4}\rightarrow 0$ everywhere.} to some background solution $(\phi_{0},g_{0})$. We will assume that $\phi$ is unconstrained, since we want to reach a decoupled quantum theory, which necessarily contains off-shell $\phi$ configurations.  We will write the full field configuration for $g$ as
\begin{equation}
	\label{metricExpansion}
	g_{\mu\nu} = \left(g^{0}_{\mu\nu} + \Delta g_{\mu\nu}\right) + \delta g_{\mu\nu}
\end{equation}
where $\Delta g_{\mu\nu}$  is chosen such that $g^{0}_{\mu\nu} + \Delta g_{\mu\nu}$ satisfy Einstein's equations with matter configuration $\phi$, while $\delta g_{\mu\nu}$ represents the remaining fluctuation.  We now integrate out the metric.  By the saddle point approximation, i.e. Einstein's equation, $\Delta g_{\mu\nu}$ is of order $M_{p}^{-2}$ and so vanishes as $M_{p}\rightarrow \infty$.   The one loop correction around the saddle is automatically absorbed into a renormalization of the matter couplings.  Finally, the (renormalized) action evaluated on the background is just the decoupled matter result.  An expansion in $\Delta g_{\mu\nu}$ thus looks schematically like:
\beq
	\label{sgexp}
	S_{g} &=&S_{EH+GHY}^{ren}(g_{0}) + M_{p}^{2} \left( a_{1} \Delta g + a_{2} \Delta g^{2} + a_{3} \Delta g^{3} + \cdots\right) \\ \nonumber
	&+& S^{ren}_{matter}(\phi)+ \left( b_{1} \Delta g + b_{2} \Delta g^{2} + \cdots\right)
\eeq
where the $a_{i}$ series comes form expanding the gravitational action and the $b_{i}$ series comes from the matter action.  Now, Einstein's equations imply `$b \sim M_{p}^{2} a$' and so simple power counting tells us that the only terms that can contribute in the decoupling limit are the linear variations.  Moreover, these linear terms contribute at the same order as $S^{ren}_{matter}$.  Thus, we just need to expand the gravitational action~\eqref{gAction} to first order in the metric deviation~\eqref{metricExpansion}.  One finds:
\beq
	\label{decoupleCond}
	\delta^{(1)} S_g &=& \frac{1}{2}\int_{\mathcal{M}} d^{4}x\sqrt{-g}\left(T_{\mu\nu}-M_{p}^{2}G^{\mu\nu}\right)\Delta g_{\mu\nu} \nonumber \\ 
	&+& \frac{C_{GHY} M_p^2}{2}\int_{\partial\mathcal{M}} d^3x\sqrt{-h}\left(h^{\mu\nu}(\mathcal{K}-\mathcal{K}_{0}) - \mathcal{K}^{\mu\nu}\right)\Delta g_{\mu\nu} \nonumber \\ 
	&+&\frac{(1-C_{GHY})M_{p}^{2}}{2}\int_{\partial\mathcal{M}} d^{3}x \sqrt{-h}\; n^{\mu}\left( \nabla^\nu \Delta g_{\mu\nu} - g^{\nu\lambda} \nabla_{\mu} \Delta g_{\nu\lambda}\right) \\ \nonumber
\eeq

The first term vanishes by the equations of motion.  The second term vanishes if we choose Dirichlet boundary conditions $\Delta g_{\mu\nu} = 0$, and the last term vanishes provided we choose $C_{GHY} = 1$.  Thus, in this limit, the full action becomes $S_{g} = S^{ren}_{EH+GHY}(g_{0}) + S^{ren}_{matter}(\phi)$, which is of the desired form~\eqref{dAction} with $S_{EH+GHY}^{ren}$ playing the role of the constant term. Note that $C_{GHY} = 1$ is just the standard boundary term, which is normally motivated by the desire to formulate a well-defined variational problem.  Here, we see that the prescription is, perhaps not surprisingly, also useful for decoupling gravity in the $M_{p}\rightarrow \infty$ limit.

There are, however, a few points to keep in mind.  First, the expansion in~\eqref{sgexp} is not guaranteed to converge unless all relevant matter energy densities are bounded as $M_{p}\rightarrow \infty$.  Secondly, the prescription given above is not unique\footnote{See, for instance the discussions in \cite{Compere:2008us,Detournay:2014fva,Castro:2017mfj}.} and we may wish to choose other values of $C_{GHY}$ for other possible boundary conditions.  Finally, the argument presented above may fail in the presence of horizons / singularities since these introduce a new component to $\partial\mathcal{M}$ whereas the GHY term is typically only evaluated at the asymptotic boundary.\footnote{It is interesting to note that the Schwarzchild black hole of mass $M$ suffers from a form of `non-decoupling'. The decoupled field theory configuration corresponds to a point particle with action $\beta M$.   However, the gravitational action is $S_{g} = \beta M - S_{BH}$, the difference just being made up by the entropy induced by the horizon.  The decoupling we are interested in is related to the action of a probe in this background, rather than the background itself.}  This is the sort of caveat that will be relevant for a proper understanding of the HT instanton.

Although the discussion in this section is general enough to apply to an arbitrary off-shell configuration of $\phi$, in the examples below we will focus on cases where $\left(\phi,g\right)$ as well as $\left(\phi_0,g_0\right)$ are solutions to the full equations of motion.  Our particular interest is in the case where  $\left(\phi,g\right)$ corresponds to a tunneling solution, which we will denote as $\left(\phi_{b},g_{b}\right)$, and $\left(\phi_0,g_0\right)$ corresponds to a non-tunneling solution, denoted $\left(\phi_{i}, g_{i}\right)$.  In this case, $\delta S_{g}$ has a natural interpretation in terms of the bounce action, as explained further below.

\section{Decoupling in Euclidean Instantons}
\label{ep}

We would now like to see how the general argument for decoupling presented above fares for Euclidean tunneling calculations.  The canonical example of tunneling in a gravitational system was worked out by Coleman and de Luccia (CdL) in~\cite{Coleman:1980aw}, which we review first.  In this case, the action reduces to the form~\eqref{dAction}, indicating a successful decoupling.  We will show, however, that in the two field example related to axion monodromy, decoupling fails.

\subsection{Coleman de Luccia bounce}

In the example studied by Coleman and de Luccia, one considers a single scalar with potential $V\left(\phi\right)$ tunneling from one metastable minimum to another.  We will label these two vacua as the `false' and `true' vacua with potentials $V_{false}$ and $V_{true}$, respectively.  In the Euclidean prescription, the tunneling rate is computed by finding the action of a Euclidean solution that starts in the false vacuum `far away' and ends in the true vacuum at the origin.  One then subtracts off the non-tunneling action and computes the tunneling rate, $\Gamma$.  Labeling the tunneling and non-tunneling solutions as $\phi_b$ and $\phi_i$ respectively, we have
\begin{gather}
	\label{bng}
	\Gamma \sim e^{-B} \\ \nonumber
	B = S\left[\phi_{b}\right] - S\left[\phi_{i}\right]
\end{gather}

The key feature when adding gravity is that the energy density closes the universe, so that one ends up with solutions that are topologically $S^{4}$.  We thus require that the tunneling solution be in the false vacuum at the `north' pole and roll to the true vacuum at the `south' pole.  In the limit of a thin potential barrier  between two nearby vacua, the potential may be completely characterized by $\Delta V = V_{false} - V_{true}$ and an effective tension, $\sigma$.  In the limit where $M_{p} \rightarrow \infty$, the membrane mediating the tunneling approaches the south pole and we may write the tunneling action computed in~\cite{Coleman:1980aw} in the form
\be
	S_{g}\left[\phi_b\right] = \frac{27 \pi^{2} \sigma^{4}}{2 \Delta V^{3}} - \frac{24\pi^2 M_{p}^{4}}{V_{false}} + \mathcal{O}\left(M_{p}^{-2}\right)
\ee
The first term is just the result from the decoupled field theory and the second is a constant term corresponding to the background `no-tunneling' solution $S_g\left[\phi_i\right]$.  This is therefore of the desired form~\eqref{dAction}.  The success of decoupling in this case may be attributed to the fact that regular boundary conditions are maintained at the north pole due to a careful fine-tuning of initial conditions  at the south pole by the overshoot-undershoot procedure~\cite{Weinberg:2012pjx}. This also guarantees a small field excursion.  In the limit $M_{p} \rightarrow \infty$, the tunneling process takes place at an angular deviation of order $\sigma \sqrt{V_{false}}/(\Delta V M_{p})$ away from the south pole, which effectively looks like flat space.

\subsection{Hawking-Turok instanton}

Let us compare this with the Hawking-Turok (HT) instanton~\cite{Hawking:1998bn}.  As described by Hawking and Turok, this is an $SO\left(4\right)$ symmetric solution to the equations of motion for a Euclidean theory
\begin{equation}
	\label{HTgaction}
	S_g = \int_\mathcal{M} d^4x\sqrt{g}\left(-\frac{M_p^2}{2}\mathcal{R} + \frac{1}{2}\left(\partial\phi\right)^2 + V\left(\phi\right)\right) - C_{GHY}M_p^2\int_{\partial\mathcal{M}} d^3x\sqrt{h}\left(\mathcal{K} - \mathcal{K}_0\right)
\end{equation}
with the requirements that the potential $V\left(\phi\right)$ have a single local minimum at $\phi=0$ and $V\rightarrow \infty$ asymptotically.  One takes the $SO(4)$ symmetric ansatz
\begin{equation}
	\label{HTansatz}
	ds^2 = dr^2 + a\left(r\right)^2d\Omega_3^2 \quad\quad \phi = \phi\left(r\right)
\end{equation}
with boundary conditions chosen as
\begin{equation}
	\label{HTboundcond}
	\begin{aligned}
		a\left(0\right) &= 0 &\quad\quad \phi\left(0\right) &= \phi_{0} \\
		a'\left(0\right) &= 1 &\quad\quad \phi'\left(0\right) &= 0
	\end{aligned}
\end{equation}
These conditions are needed in order to ensure a regular configuration at the origin, or `south pole'.  The equations of motion are
\begin{align}
	\ddot{\phi} + \frac{3\dot{a}}{a}\dot{\phi} - V'\left(\phi\right) &= 0 \\
	3M_p^2\ddot{a} + a\left(\dot{\phi} + V\left(\phi\right)\right) &= 0
\end{align}

\begin{figure}[t]
  \centering
  \includegraphics[width=0.5\textwidth]{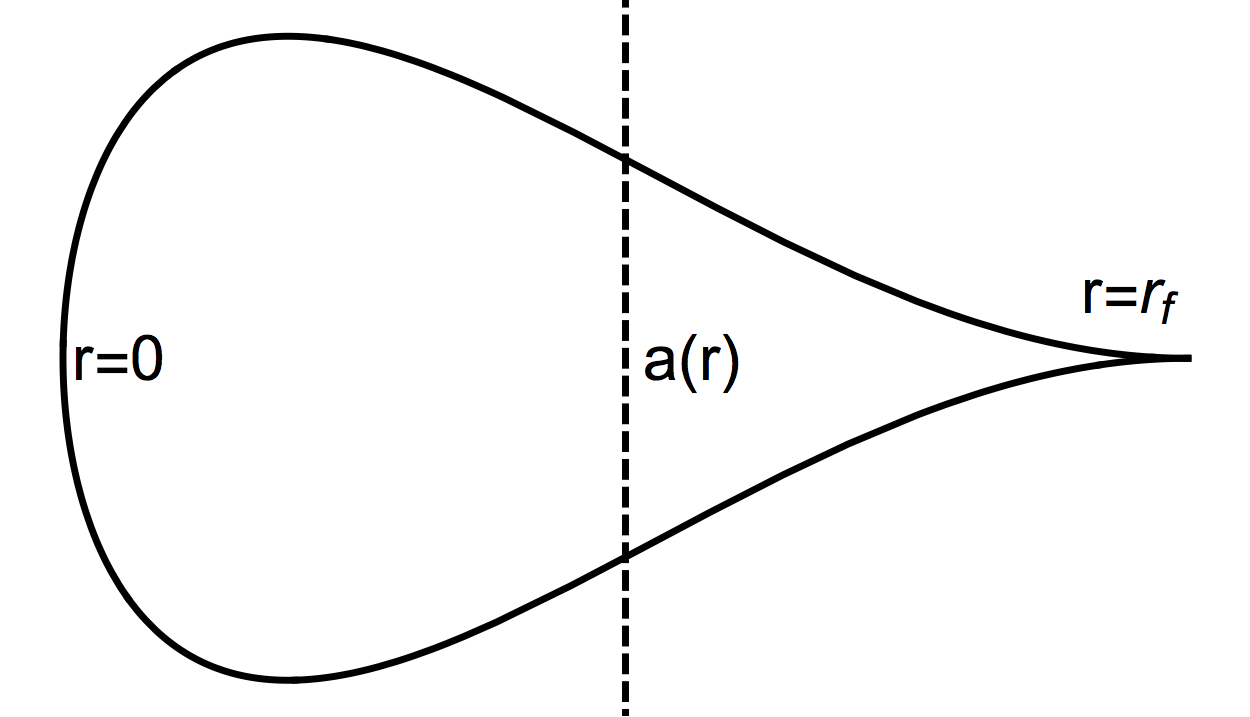}
  \caption{The geometry of the Hawking-Turok instanton.  Only the profile in the angular coordinate $r$ is shown.  The spacetime and scalar field are singular at $r=r_f$.}
  \label{fig:HTgeometry}
\end{figure}

Non-trivial solutions to these equations can only be obtained numerically even for simple potentials. However, on a qualitative level, we may think of the evolution of $\phi$ as moving in an inverted potential.   Since the inverted potential is unbounded from below, one finds that $\phi$ generically blows up. The increasing energy density then causes the geometry to close on itself.  Thus the geometry is topologically $S^4$, $r$ is an angular coordinate along the sphere, and there is a `north pole' at $r_f$ such that $a\left(r_f\right) = 0$ where the spacetime ends (see figure~\ref{fig:HTgeometry}).  At $r_f$ both the scalar and the geometry become singular with asymptotic behavior
\begin{align}
	\label{HTphiasymp}
	\phi &\sim M_p\left(\pm\sqrt{\frac{2}{3}}\ln\left(r_f-r\right) + \beta + \cdots\right) \\
	a &\sim \frac{\alpha}{M_p^{2/3}}\left(r_f-r\right)^{1/3} + \cdots
	\label{HTaasymp}
\end{align}
where $\alpha$ and $\beta$ are parameters that depend on the initial value $\phi_0$ and details of the potential.

The authors of~\cite{Hawking:1998bn} chose to set $C_{GHY} = 0$, which could be motivated on the grounds that the HT solutions are compact.  However, given the presence of the singularity, one  has the freedom to remove the singular point and treat it as a boundary, in which case the boundary term gives a non-zero contribution.  Furthermore, various other values have been proposed for $C_{GHY}$~\cite{Vilenkin:1998pp,Garriga:1998tm} based on specific choices of UV completion. We will leave it unfixed for the time being.

Let us now look at the variation of the action as we go between different solutions in the family labeled by the parameter $\phi_0$.  In general this is simply a boundary term since the bulk term vanishes by the equations of motion.  Taking the variation and substituting in the asymptotic solutions~(\ref{HTphiasymp}), (\ref{HTaasymp}) we find 
\begin{align}
	\label{dSg}
	\delta S_g&=2\pi^2 \lim_{r\to r_f}\left(3 M_p^2\left(a^2\delta \dot{a}-C_{GHY}\delta(a^2\dot{a})\right)-a^3 \dot{\phi}\delta \phi\right)\\
	&=2\pi^2\left((3C_{GHY}-1)\alpha^2\delta\alpha-\sqrt{\frac{2}{3}}\alpha^3\delta\beta\right)
	\label{dSg2}
\end{align}
where the first term comes from the variation of the metric and the second comes from varying the matter field.  Now, in order to calculate the action from here we would integrate along our family of solutions.
\begin{equation}
	\label{HTintaction}
	S_g = \frac{2\pi^2}{3}\left(3C_{GHY}-1\right)\alpha^3 - \sqrt{\frac{8}{3}}\pi^2\int \alpha\left(\beta\right)^3\; d\beta
\end{equation}
Thus, the action is determined once we know $\alpha$ and $\beta$ as functions of $\phi_{0}$.

Naively one might think that the first term in~\eqref{dSg2}, coming from the first order variation with respect to the metric, should be identical to~\eqref{decoupleCond}. After all, \eqref{decoupleCond} was supposed to be a general result.  Indeed, the metric at the boundary always satisfies $a=0$, thus implying $\Delta g_{\mu\nu} = 0$, and so the remaining variation should vanish when $C_{GHY} = 1$.   However, this is not the case for equation \eqref{dSg2}.  What is going wrong here?  The answer is that~\eqref{decoupleCond} was derived under the assumption of non-singular boundary conditions.  Consider the gravitational contribution to~\eqref{dSg}
\begin{align}
	\label{DecouplingFail}
	\lim_{r\to r_f}\left(a^2\delta\dot{a}-C_{GHY}\delta(a^2\dot{a})\right)=\lim_{r\to r_f}\left(a^2\delta\dot{a}-C_{GHY}(2a\dot{a}\delta a+a^2\delta\dot{a})\right)
\end{align}
For a regular geometry, $a\dot{a}\delta a$ vanishes as $r\to r_f$, and so the entire expression vanishes with the choice $C_{GHY}=1$. However, in our case the singularity of $\dot{a}$ causes $a\dot{a}\delta a$ to be finite at the boundary. This gives us the expression \eqref{dSg2}. The subtle change in the variation of $S_g$ is the essential reason decoupling fails for HT type singularities.

The choice $C_{GHY}=1/3$ appears to be special since this eliminates the first term in~\eqref{dSg2}.  In fact, this choice has been proposed for HT singularities for orthogonal reasons. Garriga regularized the singularity by wrapping it with a membrane and found $C_{GHY}=1/3$ in~\cite{Garriga:1998tm}. It has also been argued~\cite{Garriga:1998ri, Hashimoto:2000zk} that the singularity could be lifted to a 5d `bubble of nothing' solution~\cite{WITTEN1982481}, also resulting in $C_{GHY}=1/3$ in the 4d theory.  In fact, these lifts explain the result we get in the gravitational sector; since there is no boundary in the 5d description, the boundary term in the gravity sector must vanish.

One might naively suspect that the choice $C_{GHY} = 1/3$ leads to a decoupled theory.  However, the field values are still unbounded near the singularity and so there is no reason to trust an expansion like~\eqref{sgexp} in the first place.  Moreover, we are still left with the $\delta \beta$ term in the variation of the 4d action~\eqref{dSg}, which translates to a free parameter in the 5d lift.  We can thus learn little new from studying the lift.  The question is then whether or not the matter term in~\eqref{HTintaction} decouples when $C_{GHY} = 1/3$.  Since the fields are superplanckian and there is strong mixing between the gravity and matter sectors we have no a priori reason to suspect that decoupling works, but also no way to immediately rule this out.

To understand the situation for $C_{GHY} = 1/3$ better, we must resort to numerics.   Our goal is to compare tunneling and no-tunneling solutions, both with and without gravity.  We will moreover specialize to the `axion monodromy' potential studied in~\cite{Brown:2016nqt}.   The details are described below.

\subsection{Tunneling in the Hawking-Turok Background}

\begin{figure}[t]
  \centering
  \includegraphics[width=0.5\textwidth]{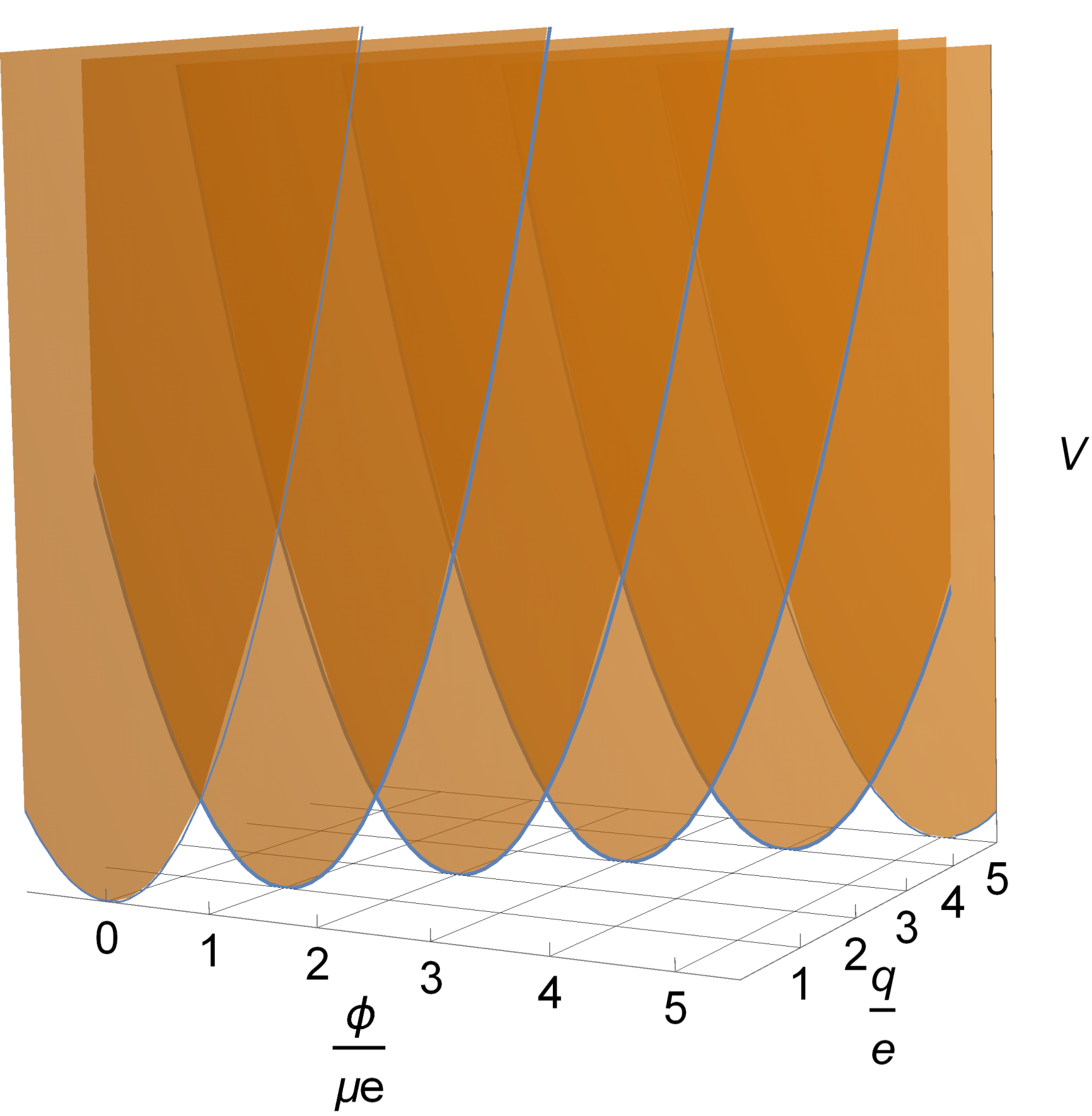}
  \caption{The potential~\eqref{NewEffPot}.  The shaded regions are above the potential while the white regions are below the potential.  The potential $V_\text{UV}$ produces deep narrow troughs in the $q$ direction where classical evolution is then allowed in the $\phi$ direction.}
  \label{fig:AMPot}
\end{figure}

The axion monodromy potential~\eqref{NewEffPot} may be thought of as a two-field potential where one field is the axion and the other provides a background.   We take the potential to be
\be
V(\phi,q) = \frac{1}{2}\mu^2\left(\phi + \frac{q}{\mu}\right)^{2} + V_\text{UV}\left(q\right)
\ee
where $V_\text{UV}$ has deep minima at integral spacing $q = n e$ (see figure~\ref{fig:AMPot}).  We want to study a process in which the scalar field rolls in the $\phi$ direction, quickly tunnels in the $q$ direction, and continues rolling.  This may be modeled by a potential of the form
\begin{equation}
	V\left(\phi\right) = \frac{1}{2}\mu^2\left(\phi - \frac{ne}{\mu}\theta\left(\bar{r}-r\right)\right)^2 + T_2^{\left(n\right)}\delta\left(\bar{r}-r\right)
\end{equation}
where $\bar{r}$ labels the radial coordinate at which tunneling occurs.

We now compute the bounce action $B_{g}$ as in~\eqref{bng} but using the gravitational action~\eqref{HTintaction}.  The non-trivial question is whether or not this approaches the field theory result computed in~\cite{Brown:2016nqt} as $M_{p}\rightarrow \infty$.  For $C_{GHY} \ne 1/3$ our general analysis suggests that the answer is in the negative, while $C_{GHY} = 1/3$ requires special attention.

\begin{figure}[t]
  \centerline{\begin{tabular}{cc} 
  \includegraphics[scale=0.55]{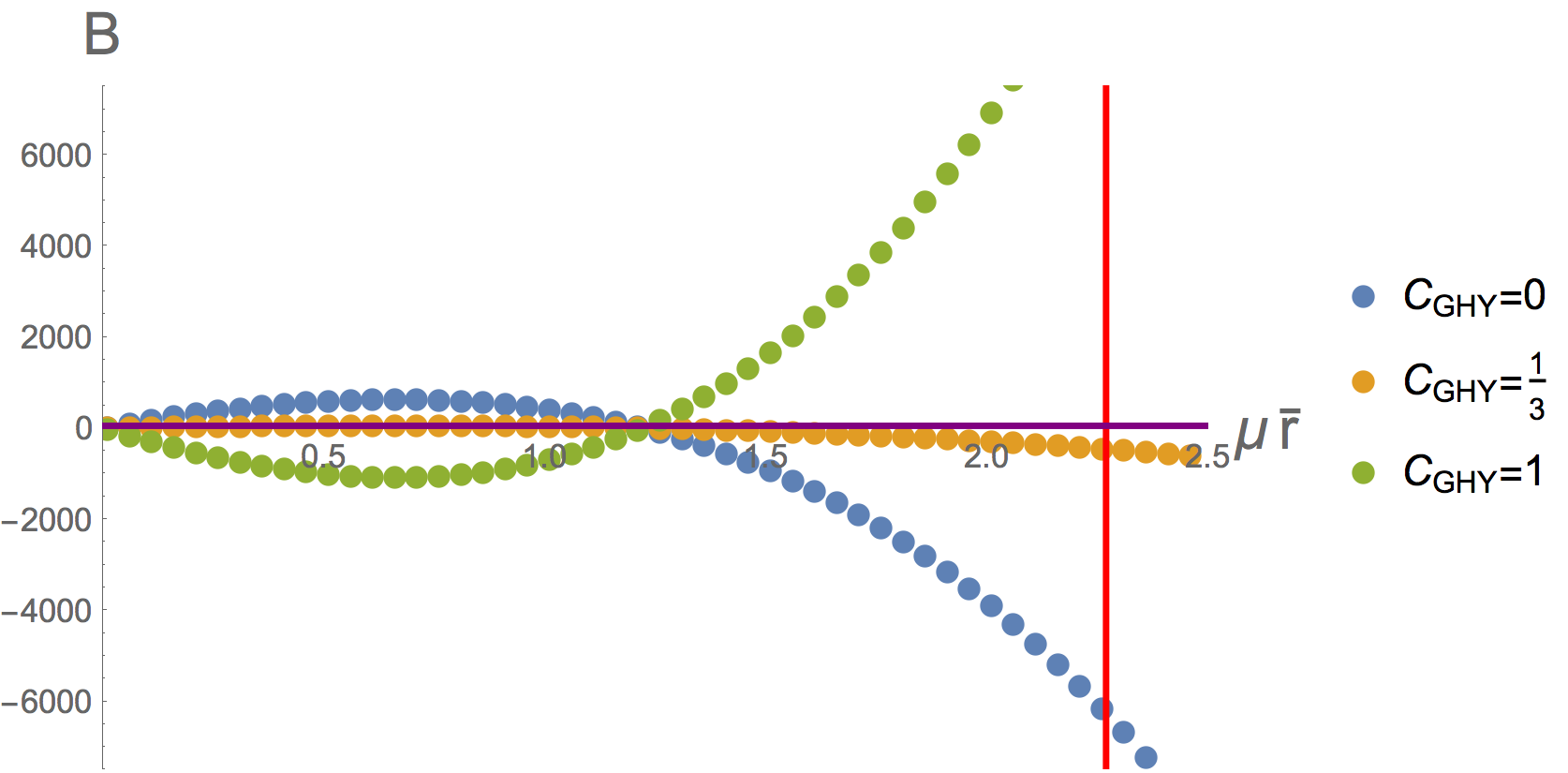} &       \includegraphics[scale=0.55]{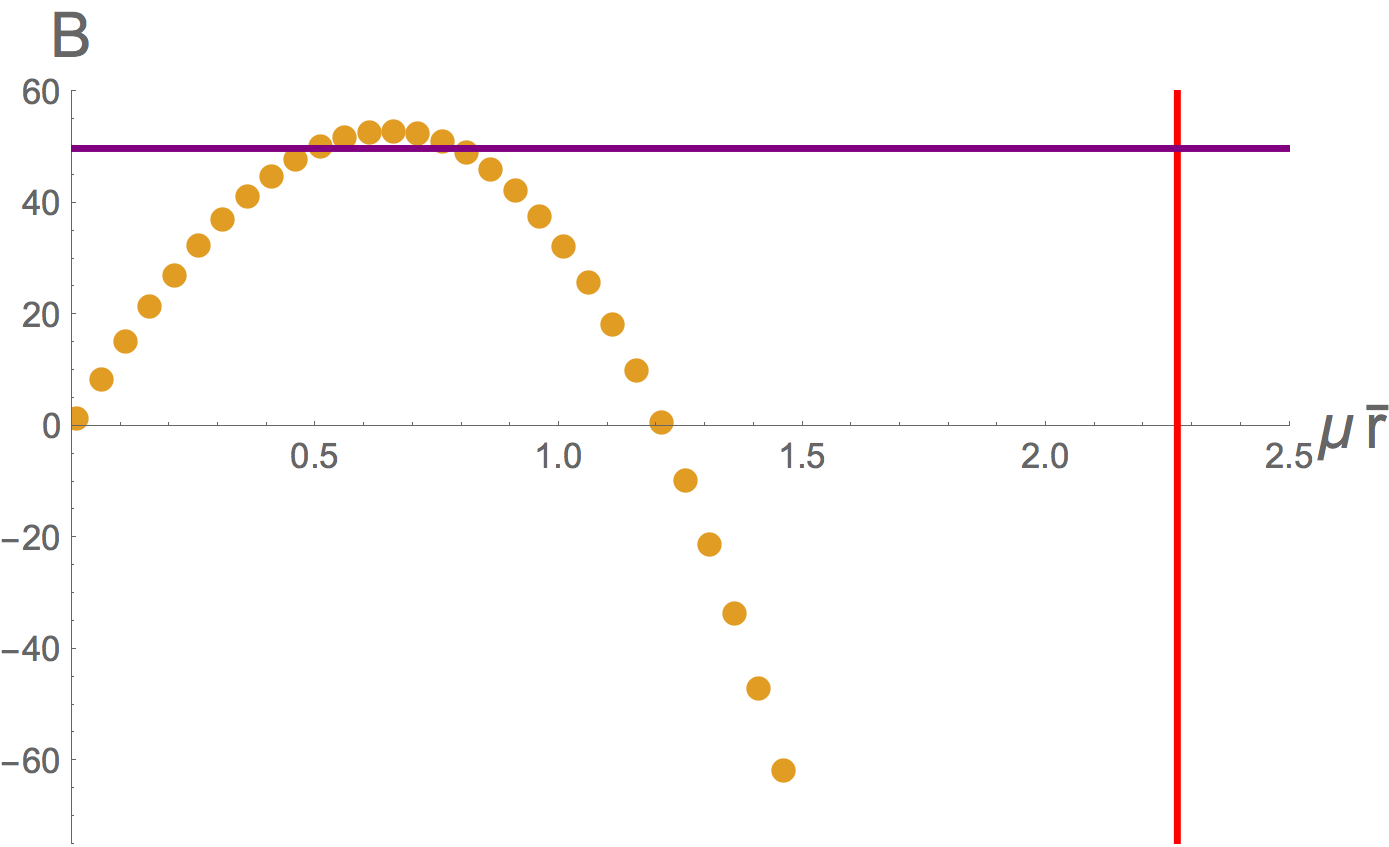} \\ 
  (a) & (b) \\ \end{tabular}}
  \caption{(a) Gravitational bounce action as a function of membrane radius for $(M_{p}/\mu)^{2} = 1000$, $e/\mu^{2} = .95$, and $T_{2}^{(1)}/\mu^{3} = .45$ for three different values of $C_{GHY}$. The purple line represents the field theory value of $B$, while the red line represents the membrane radius from the field theory calculation. (b) A close up of the plot for $C_{GHY} = 1/3$.}
  \label{results}
\end{figure}

In order to make a fair comparison between the gravitational and decoupled theories, we must choose the gravitational solution so as to reproduce the field theory solution in the vicinity of the south pole.  More precisely, we will fix the leading asymptotic behavior for the flat space solution, then require the gravitational solution to have the same behavior when zoomed in to the south pole region.  We must zoom in to a scale that is large relative to the field theory dynamics, yet small enough such that the region is flat to a good approximation.  We can then require that the gravity and field theory solutions match in this region.  In the non-gravitational theory, the general solutions are of the form
\beq
	\phi(r) = 2 A \frac{I_{1}(\mu r)}{r} + \tilde{A} \frac{K_{1}(\mu r)}{r}
\eeq
In the tunneling problem, we must match two such solutions across a domain wall while keeping the leading asymptotic behavior fixed.  The result of this analysis is~\cite{Brown:2016nqt}
\begin{align}
	\phi_{i}\left(r\right) &= 2\phi_{0}\frac{I_1\left(\mu r\right)}{\mu r} \\
	\phi_{b}\left(r\right) &=
	\begin{cases}
		2\phi_{0}\frac{I_1\left(\mu r\right)}{\mu r} + ne\mu\bar{r}^2I_2\left(\mu\bar{r}\right)\frac{K_1\left(\mu r\right)}{\mu r} &\quad\quad r > \bar{r} \\
		\frac{ne}{\mu} + \left(2\phi_{0} - ne\mu\bar{r}^2K_2\left(\mu\bar{r}\right)\right)\frac{I_1\left(\mu r\right)}{\mu r} &\quad\quad r < \bar{r}
	\end{cases}
\end{align}
From this we see that the relationship between the boundary conditions at $r=0$ is
\begin{equation}
	\label{membraneBC}
	\phi_b\left(0\right) = \phi_{0} + \frac{ne}{\mu} - \frac{1}{2}ne\mu\bar{r}^2K_2\left(\mu\bar{r}\right)
\end{equation}
where $\overline{r}$ is determined by extremizing the bounce action in the field theory system.\footnote{Of course, for the gravitational case this is just an auxiliary procedure needed to determine the correct boundary conditions.  After this step, we will also need to extremize the gravitational action with respect to the membrane radius as well.}

Our task is now to compute numerically the gravitational bounce action with boundary conditions given by~\eqref{membraneBC}.   As one can see in figure~\ref{results}, the gravitational theory generically differs dramatically from the field theory.  The case $C_{GHY}= 1/3$ comes closest to matching with field theory, though even in this case one can see that the extremized radius differs significantly from the decoupled result.  Moreover, the extremized action differs somewhat from the field theory value with a gap that increases monotonically with increasing $M_{p}$.  Thus, even in the optimal case, gravity does not decouple from matter.

Although our focus here has been on spherically symmetric tunneling processes, much of the discussion could equally well apply to the s-wave component of any process in this geometry~\cite{Brown:2016nqt}.  The membrane could be thought of as just a convenient book-keeping device to maintain regularity at the south pole in the presence of a perturbation.  One may thus conclude that the analysis presented here applies as well to general perturbations.

\section{Picard-Lefschetz: an alternate approach}
\label{pl}

Previously, one could have held the view that Euclidean methods served as a first approximation of some more rigorous approach to `dynamical tunneling' problems.  However, given the failure of these methods to recover the field theoretic limit, we claim that this is highly unlikely.  We thus seek instead an approach that guarantees agreement with field theory in the appropriate limit and admits a systematic expansion in terms of some control parameters.

Given the arguments around~\eqref{decoupleCond} it is apparent that Lorentzian amplitudes with non-singular boundaries will automatically decouple as expected.   Moreso, given generic initial conditions such as~\eqref{HTboundcond} we are not forced to consider a singular boundary as with the Euclidean HT solution.  A physically reasonable tunneling problem has regular initial conditions and the solution remains tame in the Lorentzian approach, though of course we are also free to choose sufficiently singular boundary conditions such that decoupling fails.  In contrast, Euclidean methods often translate tame initial conditions into an unbounded solution.

However, the standard problem with the Lorentzian approach is that one must deal with highly oscillatory integrals, and it is difficult to explicitly see the exponentially decaying behavior that one expects in tunneling problems.\footnote{Although see~\cite{Andreassen:2016cff}.}  Fortunately, this problem may be surmounted with Picard-Lefschetz (PL) theory~\cite{Witten:2010cx}.  PL has already found applications for the sign problem~\cite{Cristoforetti:2012su}, resurgence~\cite{Cherman:2014ofa,Behtash:2015loa}, and quantum cosmology~\cite{Feldbrugge:2017kzv,Feldbrugge:2017fcc,DiazDorronsoro:2017hti,Feldbrugge:2017mbc}.  While Euclideanization is based on complexification of the time coordinate and contour deformation of the action, PL instead complexifies the fields themselves.  Rather than perform the path integral over a real slice in field space, PL instructs us to choose a particular sum of complex slices known as Lefschetz thimbles and denoted by $\mathcal{J}_\sigma$.  The thimbles are chosen such that the original path integral is formally unchanged via an extension of Cauchy's Integral Theorem.\footnote{Of course, this is merely formal, as the proof does not immediately extend to the infinite dimensional path integral.}  The desired effect of the deformation is the same---the path integral along each thimble is exponentially damped instead of oscillatory.  In contrast to Euclideanization, however, certain complex phases may remain in the action.

As a result of the complexification of fields, PL has complex saddle points that contribute to the semiclassical expansion in addition to standard real saddle points.  It has been shown~\cite{Behtash:2015kva,Behtash:2015zha,Behtash:2015loa} that these additional complex saddle points are, in fact, necessary to correctly preserve supersymmetry in supersymmetric QM models and are the underlying exact saddle points represented by real multi instantons, which only approximately satisfy real equations of motion.

To define the PL procedure more precisely we will start with a one dimensional integral as a warm-up
\begin{equation}
	\label{PLint}
	I\left(\lambda\right) = \int dz \; e^{iS\left(z\right)/\lambda}
\end{equation}
For real coordinate $z$ and function $S\left(z\right)$, such an oscillatory integral is difficult to compute, even numerically.  Instead, the strategy of PL is to continue both $z$ and $S\left(z\right)$ into the complex plane.  Defining the exponent of the integrand $\mathcal{I} \equiv iS\left(z\right)/\lambda$, we can write this in terms of real and imaginary parts as $\mathcal{I} = h + iH$.  It is clear that if we find a homologous integration contour for which $\lim_{z\to\pm\infty} h \to -\infty$ and $H = \text{const}$ then the original integral~\eqref{PLint} becomes both non-oscillatory and manifestly convergent.  Our goal now is to construct such a contour.

We begin by listing the saddle points of $\mathcal{I}\left(z\right)$, which we label as $z_\sigma$.  Around each saddle point we are interested in the lines of steepest descent, treating $h$ as a Morse function.  These contours are called Lefschetz thimbles $\mathcal{J}_\sigma$, labeled according to their respective saddle points.  Generically, along a thimble $h$ is maximal at the saddle point itself, decreasing unbounded away from the saddle point in either direction.  The thimbles can be determined by solving the flow equations
\begin{equation}
	\label{flowEQ}
	\frac{dz^i}{du} = \pm g_{ij}\frac{dh}{dz^j}
\end{equation}
where $u$ parameterizes the contour.  The thimbles $\mathcal{J}_\sigma$ are determined by choosing the minus in~\eqref{flowEQ}.  Choosing instead the plus gives steepest ascent contours, labeled $\mathcal{K}_\sigma$.  All these contours must approach the saddle point as $u\to -\infty$.  In the flow equation~\eqref{flowEQ} we are free to choose an arbitrary Riemannian metric $g_{ij}$.  For our purposes we will choose the metric $ds^2 = \left|dz\right|^2$.  In this case the flow equations simplify to
\begin{align}
	\frac{dz}{du} &= \pm \frac{d\bar{\mathcal{I}}}{d\bar{z}} \\
	\frac{d\bar{z}}{du} &= \pm \frac{d\mathcal{I}}{dz}
\end{align}
Our contour is chosen to have the desired properties for $h$, but what about $H$?  It turns out that $H$, the imaginary part of $\mathcal{I}$, is necessarily constant along the thimble
\begin{equation}
	\frac{dH}{du} = \frac{1}{2i}\frac{d\left(\mathcal{I}-\bar{\mathcal{I}}\right)}{du} = \frac{1}{2i}\left(\frac{d\mathcal{I}}{dz}\frac{dz}{du} - \frac{d\bar{\mathcal{I}}}{d\bar{z}}\frac{d\bar{z}}{du}\right) = 0
\end{equation}
Thus the thimbles have exactly the right properties to make the integration along them convergent.

The original integral can then be rewritten as a sum over integrals along thimbles, called the Lefschetz decomposition
\begin{equation}
	\label{LefDecomp}
	\int_{\mathcal{C}_{\mathbb{R}}} dz \; e^{iS\left(z\right)/\lambda} = \sum_\sigma n_\sigma \int_{\mathcal{J}_\sigma} dz \; e^{iS\left(z\right)/\lambda} = \sum_\sigma n_\sigma I_\sigma\left(\lambda\right)
\end{equation}
The integer coefficients $n_\sigma = 0,\pm 1$ pick out exactly which thimbles should contribute to the integral.  We can determine these coefficients topologically.  Generically, any ascent $\mathcal{K}$ and descent $\mathcal{J}$ contours have intersection number
\begin{equation}
	\left<\mathcal{J}_\sigma,\mathcal{K}_\eta\right> = \delta_{\sigma\eta}
\end{equation}
This allows us to determine the coefficients $n_\sigma$ in terms of intersection numbers
\begin{equation}
	n_\sigma = \left<\mathcal{C}_{\mathbb{R}},\mathcal{K}_\sigma\right> \;\;\; \text{mod} \; 2
\end{equation}
since the real integration contour can be deformed along each $\mathcal{K}_\sigma$ until it coincides with the corresponding $\mathcal{J}_\sigma$.

There is a  possible complication to the Lefschetz decomposition,  due to the Stokes phenomenon.  Specifically, if the steepest descent path from one saddle point runs into another saddle point, the Lefschetz decomposition~\eqref{LefDecomp} is ambiguously defined.  There are two possible steepest descent paths: once the descent path from one saddle point hits another saddle point, the descent flow could branch in either direction along the thimble for the second saddle point.  In this scenario we say that the value of $\lambda$ lies on a Stokes ray of $I\left(\lambda\right)$.  To proceed, one should perturb the action so that the the thimble becomes unambiguous.  It is a non-trivial fact that the result will not depend on which continuation is picked out by the perturbation.

Although unproven, these ideas seem to be applicable to the path integral.  After complexification of fields, the Lefschetz decomposition describes a sum over integrals with nice convergence properties
\begin{equation}
	\int_{\mathcal{C}_\mathbb{R}}\mathcal{D}\phi \; e^{iS\left[\phi\right]} = \sum_\sigma n_\sigma \int_{\mathcal{J}_\sigma}\mathcal{D}\phi \; e^{iS\left[\phi\right]}
\end{equation}
Due to convergence along the thimbles, this formalism could be a viable alternative to Euclideanization for computing path integrals without having to change the signature of the metric.  It should also be noted that the Lefschetz decomposition is a geometric formalization of the standard semiclassical saddle point expansion of the path integral.

\subsection{Examples of tunneling and gravitational decoupling with PL}

We would now like to present  an example of successful decoupling in a gravitational tunneling process using Picard-Lefschetz as a proof of principle.  The model we choose is based on an exactly solvable scalar-gravity model~\cite{Garay:1990re}.  The general action takes the form
\begin{equation}
	\label{SGaction}
	S_g = \int d^4x\sqrt{-g}\left(\frac{M_p^2}{2}\mathcal{R} - \frac{1}{2}\left(\partial\phi\right)^2 - \sqrt{\frac{3}{2}}\alpha\cosh\left(\sqrt{\frac{2}{3}}\frac{\phi}{M_p}\right)\right)
 + S_{GHY}
\end{equation}
In particular, we may consider the case where $\alpha$ is negative and compute the amplitude to tunnel through the inverted $\cosh$ potential to the other side.  This is somewhat ad hoc in comparison to our original goal of studying two field tunneling or even single field vacuum decay, but nevertheless serves to illustrate the utility of PL techniques in principle.  We will work with a minisuperspace ansatz
\begin{align}
	\label{metAnsatz}
	ds^2 &= -\frac{N^2}{a\left(t\right)^2}dt^2 + a\left(t\right)^2d\Omega_3^2 \\
	\phi &= \phi\left(t\right)
\end{align}
where $d\Omega_3$ is a homogeneous isotropic 3-metric with curvature $k=1$.  In this model we will look for complex solutions to the fields including the lapse function $N$.  When $N$ is imaginary the solution will be effectively equivalent to a solution of the Euclidean equations of motion.

Our path integral takes the form
\begin{equation}
	Z = \int \mathcal{D}N\mathcal{D}a\mathcal{D}\phi\; e^{iS_g}
\end{equation}
By choosing the gauge $\dot{N}=0$, the path integral in $N$ simplifies to a one dimensional integral~\cite{Halliwell:1988wc}
\begin{equation}
	Z = \int_\mathcal{C_\mathbb{R}} dN \int\mathcal{D}a\mathcal{D}\phi\; e^{iS_g}
	\label{SGpathint}
\end{equation}
Here our contour $\mathcal{C}_\mathbb{R}$ is the positive real $N$-axis.\footnote{This corresponds to computing a propagator, i.e. a Green's function of the Wheeler-DeWitt equation~\cite{Halliwell:1988wc}. Choosing the entire real line $\mathcal{C}_\mathbb{R}=\mathbb{R}$ would instead give a solution to the Wheeler-DeWitt equation.  Which contour to choose in the context of the Hartle-Hawking wavefunction has been debated in~\cite{DiazDorronsoro:2017hti,Feldbrugge:2017mbc}.  However, since we are interested in transition amplitudes the contour giving the propagator is the correct choice for our purposes.} We also rescale $N$ such that the time interval for which this path integral is defined is simply $t\in\left[0,1\right]$.  In this gauge a physical time interval can be read off from the metric
\begin{equation}
	\Delta\tau = \int_0^1 dt\;\frac{N}{a}
\end{equation}
where the physical time $\tau$ is related to the coordinate time $t$ as
\begin{equation}
	\label{TimeReparam}
	d\tau = \frac{N}{a}dt
\end{equation}
Without loss of generality, we take $\tau=0$ when $t=0$.

Now we can write the action as
\begin{equation}
	S_g = 2\pi^2\int_0^1 dt \left(3M_p^2\left(N - \frac{a^2\dot{a}^2}{N}\right) + \frac{a^4}{2N}\dot{\phi}^2 - \sqrt{\frac{3}{2}}\alpha Na^2\cosh\left(\sqrt{\frac{2}{3}}\frac{\phi}{M_p}\right)\right)
\end{equation}
One can make the following field redefinitions
\begin{align}
	\label{xydef}
	x &= \sqrt{\frac{3}{2}}M_pa^2\cosh\left(\sqrt{\frac{2}{3}}\frac{\phi}{M_p}\right) \\
	y &= \sqrt{\frac{3}{2}}M_pa^2\sinh\left(\sqrt{\frac{2}{3}}\frac{\phi}{M_p}\right)
\end{align}
In these variables, the action simplifies considerably 
\begin{equation}
	S_g = 2\pi^2 \int dt \left(3M_p^2N - \frac{1}{2N}\dot{x}^2 + \frac{1}{2N}\dot{y}^2 - \frac{N\alpha}{M_p}x\right)
\end{equation}
Importantly, the equations of motion for $x$ and $y$ are now decoupled.  The equations of motion are
\begin{gather}
	\ddot{x} - \frac{N^2\alpha}{M_p} = 0 \quad\quad\quad \ddot{y} = 0 \\
	3M_p^2 + \frac{1}{2N^2}\dot{x}^2 - \frac{1}{2N^2}\dot{y}^2 - \frac{\alpha}{M_p}x = 0
	\label{Noem}
\end{gather}
We also impose the boundary conditions $a(0) = a_{1},\,a(1) = a_{2},\,\phi(0) = \phi_{1},\,\phi(1) = \phi_{2}$, and likewise for $x$ and $y$.  The general solution to the equations of motion are then
\begin{gather}
	\label{solPL}
	y = y_1 + \left(y_2 - y_1\right)t \\
	x = x_1 + \left(x_2 - x_1 - \frac{N^2\alpha}{2M_p}\right)t + \frac{N^2\alpha}{2M_p} t^2 \\
	N = \pm\frac{\sqrt{2}}{\alpha}\sqrt{\alpha M_p\left(x_2+x_1\right) - 6M_p^4 \pm \sqrt{36M_p^8 - 12M_p^5\alpha\left(x_2+x_1\right) + M_p^2\alpha^2\left(4x_1x_2 + \left(y_2-y_1\right)^2\right)}}
\end{gather}
It is now trivial to determine the action for any initial conditions.  Note that $N$ has four solutions which we will label as $N_{\pm\pm}$.  The one non-trivial input from PL theory is to tell us which  saddle points contribute in the Lefschetz decomposition.

\subsubsection{Lorentzian tunneling with PL}

\begin{figure}[t]
  \centering
  \includegraphics[width=0.5\textwidth]{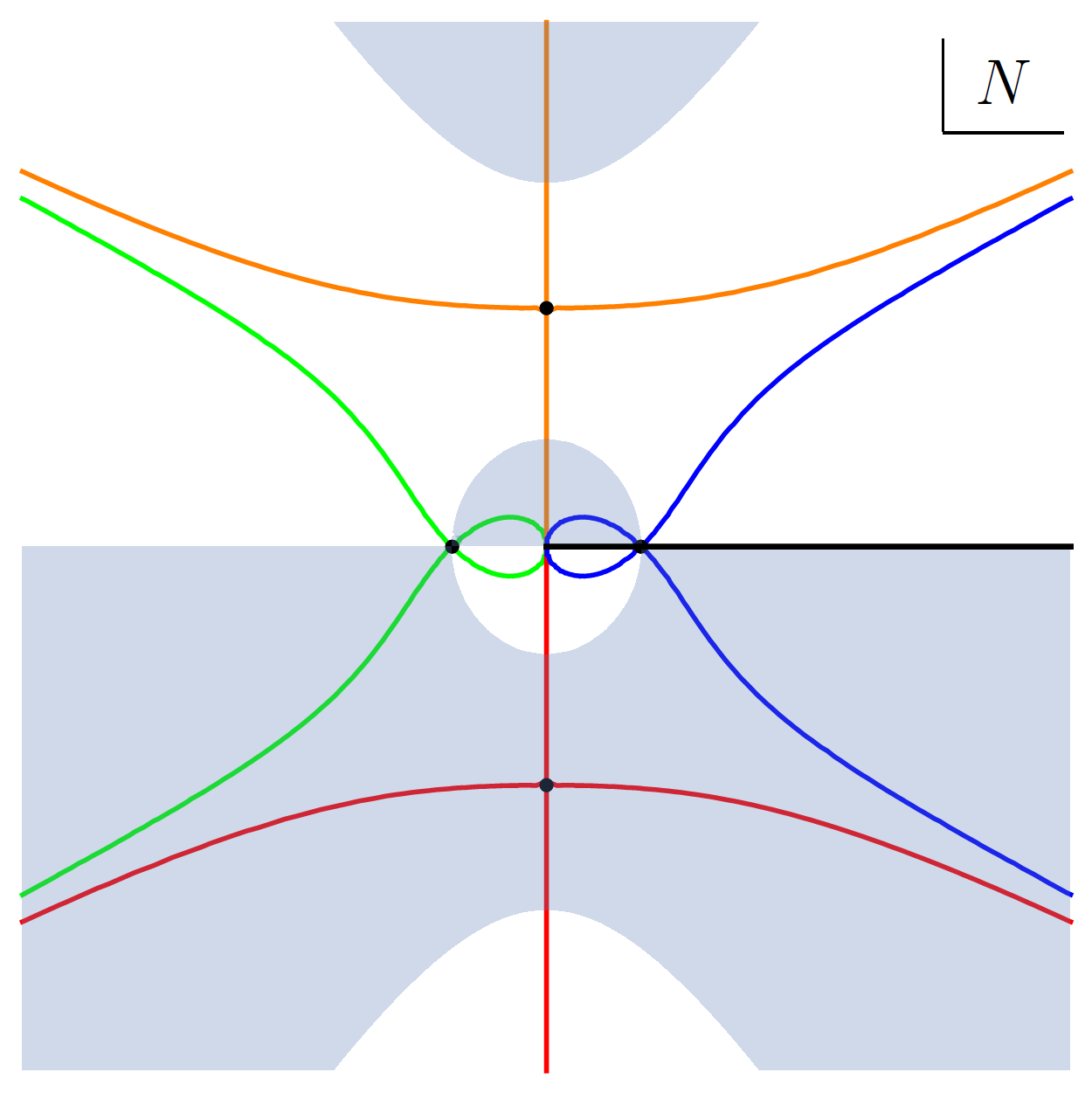}
  \caption{The solutions for the boundary condition $a_1=a_2$.  The flow lines (and corresponding saddle points) correspond to the solutions $N_{\pm\pm}$ as follows: ($++$:blue),($+-$:orange),($-+$:green),($--$:red).  The shaded region has $\text{Re}\left(iS_g\right)>0$.  The black line represents the integration contour for a Lorentzian path integral over the interval $N\in\left(0,\infty\right)$.}
  \label{fig:FlatPLgraph}
\end{figure}

For an explicit test we need only specify the boundary conditions.  As discussed earlier, decoupling is guaranteed to work as long as the boundaries are chosen to be non-singular.  For simplicity we will choose $a_1 = a_2\ne 0$ and study transitions between different values of~$\phi$.   Given this, we need only to plug the solutions back into the action to find the leading contribution.

We would like to expand the action in powers of $\frac{1}{M_p}$ to make the decoupling limit apparent. We must be careful about what is kept fixed in this limit.  If we  keep $a_1=a_2=\text{const}$ in this limit, then the physical time interval $\Delta \tau$ generically goes to zero.  Instead, we choose to hold the physical time interval constant in order to reproduce a finite time process in the decoupled theory.  As a result, the scale factor $a_1=a_2 \sim M_p$.  This, however, is the behavior we expect.  Since the scale factor behaves like the radius of curvature for a spatial slice in this model it must diverge in the decoupling limit to reproduce a Minkowski spacetime.  Alternatively, we may think of `zooming in' on a small local patch as equivalent to letting the scale factor blow up.

Now, with the PL prescription beginning with a Lorentzian contour,  only the upper blue branch of figure~\ref{fig:FlatPLgraph} corresponding to the $N_{++}$ solution contributes to the path integral.\footnote{In order to make this equivalence exact one must check that the extra contour contributions from the origin and infinity that close the deformation region vanish.  Fortunately, this has already been proven in~\cite{Feldbrugge:2017kzv} for cases such as the one presented here.}   The action and solutions may be expanded as
\begin{align}
	a\left(\tau\right) &= a_1 + \mathcal{O}\left(\frac{1}{M_p}\right) \\
	\label{PLresult}
	\phi\left(\tau\right) &= \phi_1 + \frac{\phi_2-\phi_1}{\Delta\tau}\tau + \mathcal{O}\left(\frac{1}{M_p}\right) \\
	\label{intgAction}
	\frac{S_g}{\mathcal{V}} &= \frac{\left(\phi_2-\phi_1\right)^2}{2\Delta\tau^2} + \mathcal{O}\left(\frac{1}{M_p}\right)
\end{align}
where $\mathcal{V} = \int d^4x \sqrt{-g}$ is the spacetime volume.  Now we want to compare this against the solutions and integrated action for the corresponding decoupled theory.  Again, this will work out in a trivial manner.  Since the potential $V \propto \cosh\left(\sqrt{\frac{2}{3}} \frac{\phi}{M_{p}}\right) \rightarrow 1$ in the decoupling limit, we arrive at a free theory.   It is then easily seen that the result agrees with~\eqref{PLresult} and~\eqref{intgAction}.

When $\alpha < 0$ the potential is an inverted $\cosh$ and a tunneling amplitude corresponds to having $\phi_1$ and $\phi_2$ on opposite sides of the maximum.  Although the theory is free in the decoupling limit, the tunneling amplitude is nevertheless well defined for any finite $M_p$ and gives a non-trivial example of tunneling in a gravitational theory where we do not begin and end at a metastable vacuum.  This would have been impossible with Euclidean techniques.  The use of Euclidean techniques requires a suitable continuation back to Lorentzian signature.  In order to avoid complex momenta, this must be done along a spacetime surface where $d\phi/d\tau = 0$.  However, for the `Euclidean' solution in this model $N_{--}$, no such surface exists.

\begin{figure}[t]
  \centering
  \includegraphics[width=0.5\textwidth]{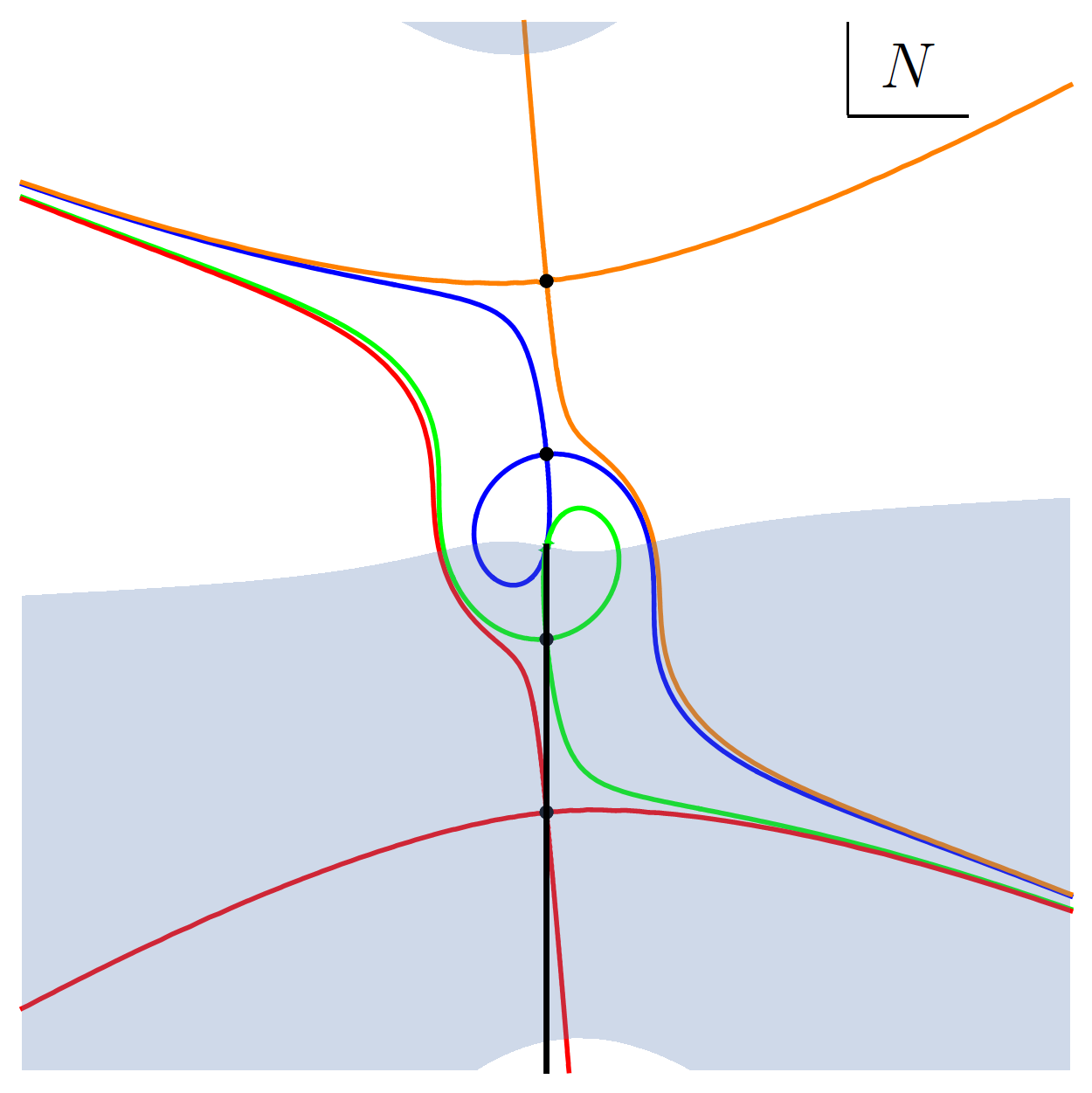}
  \caption{The solutions for the boundary condition $a_1=0$.  The flow lines (and corresponding saddle points) correspond to the solutions $N_{\pm\pm}$ as follows: ($++$:blue),($+-$:orange),($-+$:green),($--$:red).  The shaded region has $\text{Re}\left(iS_g\right)>0$.  The black line represents the integration contour for a Euclidean path integral over the interval $N\in\left(0,-i\infty\right)$.  In order to avoid a Stokes ray~\cite{Witten:2010cx} we have sent $\hbar\to 1+0.2i$.}
  \label{fig:HTPLgraph}
\end{figure}

\subsubsection{Euclidean solutions with PL}
\label{EucPL}

It is also of interest to consider solutions with imaginary $N$ within the PL framework as these may form a bridge between standard Euclidean and Lorentzian amplitudes.  For instance, if we choose $a_1 = 0$ and $a_2 > 0$, the geometry~\eqref{metAnsatz} can be viewed as a cone with $t$ the radial direction.  As seen in figure~\ref{fig:HTPLgraph}, all of the saddle points for this choice of boundary conditions lie on the imaginary $N$ axis.  Thus the geometry of the saddles automatically has Euclidean signature.  This is somewhat suggestive, since this is precisely the geometry needed to describe either pole in the Euclidean tunneling calculation.

We will begin by examining the decoupled Euclidean theory with metric $ds^{2} = dt^{2} + t^{2} d\Omega_{3}^{2}$ and boundary conditions $\phi(\epsilon)=\phi_{1}$, $\phi(1) = \phi_{2}$.  This should be analogous to the gravitational boundary condition $a_1 = 0$ and  $\phi(0) = \phi_{1}$ in the limit $\epsilon \to 0$.  However taking this limit in field theory on the flat background leads to a degeneration of the solutions for $\phi$ such that the only solutions are constants.  This is equivalent to imposing the regularity condition $\phi'(0) = 0$, which is part of the standard Euclidean prescription, though here we are thinking in terms of the Lorentzian procedure.  The field theory action density evaluated in the $\epsilon\rightarrow 0$ limit is then what we would expect for constant solutions
\begin{equation}
	\frac{S_{d,E}}{\mathcal{V}} = \sqrt{\frac{3}{2}}\alpha
\end{equation}

Now we will look at the gravitational theory~\eqref{SGaction} with the same boundary conditions as above.  As with the previous example, there are four solutions for the gravitational theory but only one for the corresponding decoupled theory.  Since the decoupled theory we are considering has Euclidean signature, we will start from a Euclidean path integral contour in the complex $N$ plane as indicated by the black line in figure~\ref{fig:HTPLgraph}.  However, as we can see in that figure, PL tells us that there are two saddle points that contribute to this path integral: $N_{-+}$ and $N_{--}$.  The action evaluated on the two solutions in the limit $M_{p}\rightarrow \infty$ then becomes\footnote{Now keeping $a_{2}$ fixed.}
\begin{align}
	\frac{S_g\left[N_{-+}\right]}{\mathcal{V}} &= \frac{12iM_p^2}{a_2^2} + \mathcal{O}\left(M_p^0\right) \\
	\frac{S_g\left[N_{--}\right]}{\mathcal{V}} &= i\sqrt{\frac{3}{2}}\alpha + \mathcal{O}\left(M_p^{-1}\right)
	\label{gActionAlt}
\end{align}
In the decoupling limit the contribution to the path integral~\eqref{SGpathint} from the thimble around the $N_{-+}$ solution becomes infinitely suppressed.  This means that the solution $N_{--}$ is the only one that contributes to the path integral and thus the action evaluated on this solution~\eqref{gActionAlt} should agree exactly with that of the decoupled theory.  We see that this is indeed the case; the $\phi_{i}$ dependence has dropped out in both theories.

It is interesting to note that for this solution the geometry generically has a conical singularity at $t = 0$ but decoupling is still successful.  However, this should not be too surprising a result.  The problem with non-decoupling for the HT instanton was not just that there was a singular point in the geometry, but rather that the singularity was specifically of the HT type~\eqref{HTaasymp}.  More precisely, the scale factor must have an asymptotic form near the singularity like $a \sim \tau^n$ where $n \leq 1/3$ in order to reproduce the failure mode of decoupling identified in~\eqref{DecouplingFail}.  However, a conical singularity with $a \sim C\tau$ for $C \neq 1$ does not suffer this problem.  As such, not all singularities are equally problematic for decoupling.  Moreover, such conical singularities are actually necessary to ensure decoupling is successful.  For the HT solution, requiring regularity at one pole~\eqref{HTboundcond} generically leads to a HT singularity at the opposite pole.  However, in the example above we see that if we take $a_2 \to 0$ while keeping $\phi_2$ finite then we produce conical singularities at both poles.  Attempting to modify parameters to force one pole to become regular then requires introduction of a HT singularity at the other causing decoupling to fail.

The key lesson here is that the solutions with conical singularities for which decoupling succeeds are precisely the solutions forbidden by the Euclidean prescription while the HT solution is the one required by that prescription.  Connecting a Euclidean solution to some real time evolution requires it to be matched onto a Lorentzian solution related by some analytic continuation.  A necessary condition for such a matching is that the time derivatives of the matter fields must go to zero along the matching surface.  This is the condition that motivates the choice of boundary conditions~\eqref{HTboundcond} for the HT solution in~\cite{Hawking:1998bn} as the Lorentzian universe considered in \cite{Hawking:1998bn} is continued from the Euclidean solution at the regular pole.  Here we see plainly the limitation of the Euclidean prescription.  The boundary conditions required by the Euclidean prescription are not the ones that lead to decoupling.  Such solutions can only be seen in a Lorentzian prescription such as PL where one is allowed more flexibility in choosing boundary conditions.  Perhaps unsurprisingly, these boundary conditions, with nonzero field velocities, are what one would expect when considering tunneling in a system with nontrivial dynamics such as tunneling during slow-roll inflation.

\subsection{Remarks}

One remaining question is how the previous results based on Euclidean techniques (i.e., instantons) fit into the PL framework.  We have seen here that PL can reproduce aspects of the Euclidean prescription, such as the regularity condition and imaginary lapse function.  This issue was also partially addressed in \cite{Cherman:2014sba}, where it was shown that a Euclidean instanton in the quartic double well system may be continuously rotated to a (singular) complex Lorentzian saddle point and that the fluctuation determinants around the Euclidean solution and corresponding Lorentzian solution match.  However, this approach begins by assuming results of the Euclidean prescription and merely translates them to the Lorentzian language.  Instead, one would like to find the instantons beginning from the PL framework so as to determine when they actually do or do not contribute.

It is intriguing to notice that a unification of Euclidean and Lorentzian solutions in a PL framework seems more approachable when coupling the theories to gravity.  To see this, consider a general gravitational metric in the ADM decomposition
\begin{equation}
	ds^2 = g_{\mu\nu}dx^\mu dx^\nu = -N^2dt^2 + 2N_idtd\vec{x}^i + g_{ij}d\vec{x}^id\vec{x}^j
\end{equation}
Since PL requires complexification of all dynamical fields, in a gravitational theory this includes complexification of the lapse function $N$.  Paths with a real lapse function correspond to a `Lorentzian' branch while paths with imaginary lapse correspond to a `Euclidean' branch. Thus, both `Euclidean' and `Lorentzian' theories manifest themselves simultaneously within a single gravitational framework.\footnote{Of course there must also be saddle points with complex $N$ which do not necessarily fall under either classification.}

This suggests a strategy for understanding the relationship between Euclidean and real-time techniques in {\it{field theories}} by studying the relationship first in {\it{gravitational theories}}.  In following this route, we cannot help but consider an Einstein-Hilbert term, since this is induced in any case by loop corrections.  However, since we are free to stipulate regular boundary conditions in the Lorentzian framework, decoupling back to the original non-gravitational field theory may be achieved.  Perhaps one can then identify the instantons from within a formalism that also allows for real time dynamics.

\section{Conclusions}
\label{conclusions}

In this work, we have studied tunneling processes motivated by axion monodromy while including gravitational backreaction.  Our main result is that in tunneling processes that involve real time dynamics (e.g., another rolling field), one generally finds that UV physics do not decouple when one uses Euclidean methods.  Moreover, our findings suggest that this non-decoupling may be attributed to a generic Hawking-Turok type singularity that emerges in theories with unbounded potentials.  We thus conclude that, for certain tunneling problems, Euclidean methods are not salvageable; i.e., the Euclidean bounce action is not the leading term in any systematic calculation scheme.  This conclusion would seem to also have implications for the no-boundary proposal.  Finally, we have argued that the Picard-Lefschetz formalism overcomes these difficulties by allowing for more general boundary conditions. As a proof of concept, we also provided a simple example of tunneling in this framework.  To our knowledge, this is the first attempt to compute tunneling in quantum field theory using the Picard-Lefschetz method.  We plan to apply this formalism to other, perhaps more realistic, setups in the future.

\subsection*{Acknowledgments}

We would like to thank Philip Argyres, Yuta Hamada, Oliver Janssen, and Aitor Landete for helpful discussions. This work is supported in part by the DOE grant DE-FG-02-95ER40896 and the Kellett Award of the University of Wisconsin.

\bibliography{PLDecouplingbib}\bibliographystyle{utphys}

\end{document}